\documentclass[preprint,showpacs,preprintnumbers,amsmath,amssymb]{revtex4}
\usepackage{amsmath,amssymb}
\usepackage{graphicx}% Include figure files
\usepackage{dcolumn}% Align table columns on decimal point
\usepackage{bm}% bold math

\begin{document}
\title{Modular synchronization in complex networks with a gauge Kuramoto model}
\author{E. Oh$^{1,2}$}
\author{C. Choi$^2$}
\author{B. Kahng$^2$}
\author{D. Kim$^2$}
\affiliation{{$^1$Bioanalysis and Biotransformation Research
Center, Korea Institute of Science and Technology, Seoul 136-791,
Korea}\\{$^2$Department of Physics and Astronomy and Center for
Theoretical Physics, Seoul National University, Seoul 151-747, Korea}}
\date{\today}
\begin{abstract}
We modify the Kuramoto model for synchronization on complex networks
by introducing a gauge term that depends on the edge betweenness centrality (BC).
The gauge term introduces additional phase difference between two vertices from 0 to $\pi$
as the BC on the edge between them increases from the minimum to the maximum in the network. When the network has a modular structure,
the model generates the phase synchronization within each module, however, not over the entire system.
Based on this feature, we can distinguish modules in complex networks,
with relatively little computational time of $\mathcal{O}(NL)$,
where $N$ and $L$ are the number of vertices and edges in the system, respectively.
We also examine the synchronization of the modified Kuramoto model
and compare it with that of the original Kuramoto model in several complex networks.
\end{abstract}
 \maketitle

Complex networks have drawn considerable attention from diverse
disciplines such as sociology, information science, physics,
biology and so on~\cite{reviews}.
Many complex networks in real world contain modules within them,
which form in a self-organized way to achieve the efficiency functionally or regionally.
Such modular systems can exhibit collective synchronized patterns
within each module, not forming the global synchronization~\cite{huang}
as can be found in the cortex of neural network~\cite{zhou} or different synchronization transition behaviors
depending on the patterns of inter-modular connections~\cite{Oh}.

In this Letter, we study the modular synchronization pattern generated from a modified Kuramoto equation (KE),
which we call the gauge KE,
\begin{equation} \frac{d\phi_i(t)}{dt}=\Omega_i-J\sum_{j=1}^N
a_{ij} \sin(\phi_i(t)-\phi_j(t)-\eta g(b_{ij})).\label{model}
\end{equation}
Here, $\phi_i$ is the phase of vertex $i$, $\Omega_i$ is the natural
frequency of vertex $i$ selected from the Gaussian distribution
$e^{-\Omega^2/2}/\sqrt{2\pi}$, $J$ is the overall coupling constant and
$a_{ij}$ is the $(i,j)$-th component of the adjacency matrix,
which is one when the vertices $i$ and $j$ are connected, and zero otherwise.
$\eta$ is a control parameter.
The extra phase term $g(b_{ij})$, we call the gauge term below, is defined as
\begin{equation} g(b_{ij})=\frac{b_{ij}-b_{\rm min}}{b_{\rm max}-b_{\rm min}}\pi,\end{equation}
where $b_{\rm min}$ and $b_{\rm max}$ are the minimum and the maximum
edge betweenness centrality (BC)~\cite{freeman} or load~\cite{load},
respectively, in the system. Here, the edge BC or load is the amount of
effective traffic passing through a given edge when every pair of vertices sends and receives a unit packet that travels along the shortest path between them.
Then the gauge term $g(b_{ij})$ is in the range
from 0 to $\pi$ depending on the BC of edge.
When $\eta = 0$, the gauge KE recovers the standard KE~\cite{kuramoto} which becomes fully synchronized when $J$ is sufficiently large.
%Its dynamics has been used in ~\cite{arenas} to reveal community structures.
The KE with the extra phase of the form $\sin(\phi_i-\phi_j-c)$ ($c=$ constant) was studied first in~\cite{sakaguchi}.
The effect of the extra phase is to destroy the synchronization. Intuitively, one expect that the BCs on intra-module links are smaller than those on inter-module.
Thus, each module can be synchronized, while the entire system is not.
Moreover, the gauge term induces an effective coupling that can be negative at the edges connecting different modules.
Due to this negative coupling, the average phase of each module may have velocity different from each other.
Using this property, the gauge KE can be used for module identification in complex networks.

The module identification in the context
of synchronization has been studied~\cite{arenas,boccaletti}.
These studies are inspired by the so-called dynamic clustering (DC)
approach that individual oscillators have different levels of
synchronization time owing to the heterogeneity of degree in network.
Since vertices within modules are densely connected, they
are synchronized more earlier than those between modules. Using this
idea, the hierarchical structure can be detected by monitoring
the temporal evolution of synchronization~\cite{arenas}. To identify the
modules, however, the information of characteristic time at each
hierarchical level is needed, which may be obtained from the
spectrum of the Laplacian matrix of the system.
Boccaletti {\it et al.,}~\cite{boccaletti} introduced another model,
in which the coupling strength of the KE depends on the BC as
$b_{ij}^{\alpha(t)}$, where $\alpha(t)$ is negative.
Thus, the coupling strengths across the module-connecting edges are weaker than those within module.
$\alpha(t)$ is then tuned to detect the modules.
In both methods, one needs to control the parameters such as time and $\alpha(t)$.
However, our method based on Eq.~(\ref{model}) with $\eta=1$ does not contain any control parameter, so that
we can identify the modules without any prerequisite information.

We begin to study the synchronization pattern generated from Eq.~(\ref{model}).
Firstly, we apply the gauge KE to an {\it ad hoc} network~\cite{newman} with a modular structure.
The network is composed of $N=128$ vertices and $L=1024$ edges.
Those vertices are grouped to four modules, each of which is of equal size.
And edges are connected with probability $p_{in}$ for pairs of nodes belonging to the same module
whereas pairs belonging to different modules have edges with probability $p_{out}$.
By controlling the parameter $p_{in}$ and $p_{out}$
we can obtain a fraction of inter-modular edges, $z_{\rm out}/\langle k \rangle$ as we want,
where $z_{\rm out}$ is the mean degree of inter-modular edges and $\langle k \rangle =2L/N$ is the mean degree.
This {\it ad hoc} network has been used as a benchmark for module identification algorithms
in previous studies~\cite{danon2005}.

\begin{figure}[t!]
\includegraphics[width=6cm, angle=-90]{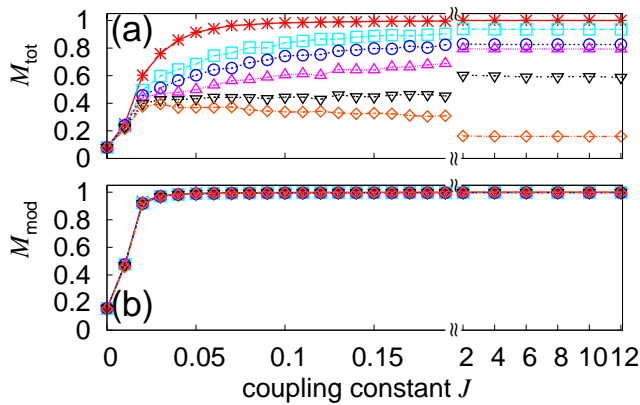}
\caption{(Color online)
The order parameter defined over the entire network (a) and within a module (b)
versus the coupling constant $J$ for the {\it ad hoc} network in case of $z_{\rm out}/\langle k \rangle=0.05$.
Data are for $\eta=0.0$, 0.6, 0.7, 0.8, 0.9 and 1.0 from the top in (a).
The same symbols are used for (b), but data for different $\eta$ collapse onto the single curve.
}\label{fig_order}
\end{figure}

\begin{figure}[t!]
\includegraphics[width=6cm, angle=-90]{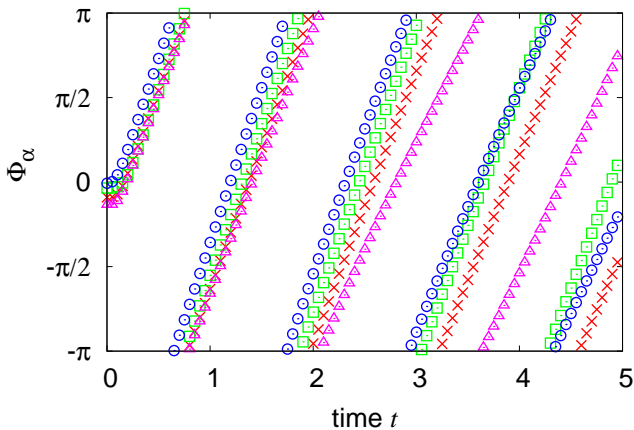}
\caption{(Color online) The time evolution of average phases of the four modules, distinguished by different symbols, for the {\it ad hoc} network with $z_{out}/\langle k \rangle=0.05$ when $\eta = 1.0$ and $J = 2.0$.
}\label{fig_ave_phase}
\end{figure}

We measure the order parameter defined as
\begin{equation}
{\cal{M}}_{\rm tot}\equiv \Bigg\langle\Bigg|
\frac{1}{N} \sum_{j=1}^N e^{i\phi_j}\Bigg|\Bigg\rangle,\label{order}
\end{equation}
where $\langle \cdots \rangle$ denotes the time and ensemble average.
The order parameter is measured in the steady state.
When $\eta=0$, the order parameter saturates to 1 for large $J$,
however, as $\eta$ is increased toward 1, it saturates at lower values as shown in Fig.~\ref{fig_order}(a).
This behavior indicates that the network is not synchronized globally.
To check if the synchronization forms within each module,
the local order parameter, defined as
${\cal{M_{\rm \alpha}}}\equiv\langle |\sum_{j=1}^{N_{\alpha}} e^{i\phi_j}/N_{\alpha}|\rangle$, is measured,
where $\alpha$ is the module index, $N_{\alpha}$ is the number of vertices within the module $\alpha$
and the sum is over vertices within the module.
We find that indeed the order parameter $\mathcal{M}_{\rm mod}$ reaches 1 for large $J$ as shown in Fig.~\ref{fig_order}(b),
indicating that the oscillators within the module are synchronized.
We examine the average phase of each module as a function of time.
As shown in Fig.~\ref{fig_ave_phase},
the modules are distinguishable by different average phases and average phase velocities.
%In this case, the global order parameter oscillates as a function of time.

\begin{figure}[t!]
\includegraphics[width=6cm, angle=-90]{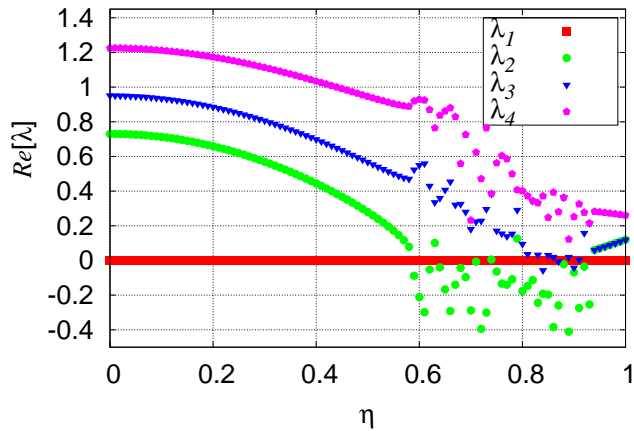}
\caption{(Color online)
The first 4 eigenvalues, $\lambda_1=0$, $\lambda_2, \lambda_3$ and $\lambda_4$, of $G_{ij}$
versus the parameter $\eta$ for the {\it ad hoc} network in case of $z_{\rm out}/\langle k \rangle = 0.05$ and $J= 2.0$.
Data beyond $\eta_c \approx 0.59$ depend sensitively on time $t$ where $\omega_{ij}$ is obtained.
}\label{fig_eigenvalue_eta}
\end{figure}

The stability of synchronization of the model (\ref{model}) is examined.
Assuming the fully synchronized state of the form $\phi_i^* = \phi_i^0 + \Omega t$, and linearizing Eq.(\ref{model}),
we get $\dot{\xi_i}(t) = -J\sum_j{G_{ij}\xi_j(t)}$
where $\xi_i(t) = \phi_i(t) - \phi_i^*$,
$G_{ij} = (\sum_k{a_{ik} \omega_{ik}}) \delta_{ij} - a_{ij} \omega_{ij}$
and $\omega_{ij} = \cos(\phi_i^0-\phi_j^0-\eta g(b_{ij}))$.
$\lambda_1=0$ is the trivial eigenvalue of $G$
and the sign of other eigenvalues determines the  stability of the fully synchronized state.
Due to the negative element of the coupling matrix $G$,
its eigenvalues can be negative, and then the Lyapunov exponent in the linear stability analysis can be as well.
In that case, the synchronization is no longer stable.
We obtain $\omega_{ij}$ from $\cos(\phi_i(t)-\phi_j(t)-\eta g(b_{ij}))$ at an arbitrary but sufficiently large $t$ and
trace out the eigenvalues for the {\it ad hoc} network having $z_{\rm out}/\langle k \rangle= 0.05$
and plot the first 3 non-zero eigenvalues versus $\eta$ in Fig.~\ref{fig_eigenvalue_eta}.
$\lambda_2$ is positive at $\eta=0$ and decreases to zero as $\eta$ increases from 0 to $\eta_c\approx0.59$.
And increasing $\eta$ further above $\eta_c$ drives the system to unstable state.
For $0 \leq \eta < \eta_c$, the order parameter ${\cal{M}}_{\rm tot}$ is almost 1 in the steady state,
whereas ${\cal{M}}_{\rm tot}$ has a smaller constant value for $\eta > \eta_c$.
In many cases, they actually oscillates in time before the time average due to disparate group velocities of the modules as shown in Fig.~\ref{fig_ave_phase}.
The curve fitting of $\lambda_2$ in the vicinity of $\eta = \eta_c$ shows $\lambda_2 \propto (\eta_c-\eta)^{1/2}$.
The square-root singularity of $\lambda_2$ near the stability edge is the signature of the saddle-node bifurcation~\cite{saddle}.

We introduce how to identify modules with the gauge KE.
To this end, we take the following steps:
\begin{itemize}
\item[$i)$] We apply the gauge KE (\ref{model}) to all oscillators with a sufficiently large coupling constant $J$.
The phases $\{\phi_i(t)\}$ of each oscillator are obtained in the steady state.
\item[$ii)$] We measure the phase similarity defined as $C_{ij}=\langle [1+\cos(\phi_i(t)-\phi_j(t))]/2\rangle$
for each connected pair of oscillators $(i,j)$.
The brackets are the average over different times, natural frequencies $\{\Omega_i \}$,
and initial random phases $\{\phi_i(0)\}$.
\item[$iii)$] From the empty state, where all edges are absent,
we add edges $(i,j)$ one by one that are chosen following the descending order of
$C_{ij}$.

Clusters after the step $iii)$ are regarded as modules.
The edges that existed originally, but not connected yet until the
step $iii)$ are regarded as inter-modular edges.
\item[$iv)$] We repeat the step $iii)$ until the modularity of the
system becomes maximum. The modularity $Q$ is defined as
\begin{align}Q =\sum_{\alpha}e_{\alpha\alpha}-a_{\alpha}^2,
\end{align}
where $a_{\alpha}=\sum_{\beta}e_{\alpha \beta}$, and $e_{\alpha \beta}$
is the fraction of edges that connect the vertices
belonging to the modules $\alpha$ and $\beta$~\cite{newman}.
\end{itemize}

To test the performance of our algorithm, we measure the mutual
information on several networks, defined as
\begin{align}
I(A,B)=\frac{-2\sum_{i=1}^{M}\sum_{j=1}^{M'}\log(\frac{N_i^j}{N_iN^j})}
{\sum_{i=1}^{M}N_i\log(\frac{N_i}{N})+\sum_{j=1}^{M'}N^j\log(\frac{N^j}{N})}
\end{align}
where $M=4$ is the number of preassigned modules and $M'$ is the
number of detected modules. $N_i^j$ is the number of vertices
belonging to the $i$-th preassigned and the $j$-th detected
modules, $N_{i}=\sum_j N_i^j$ and $N^j=\sum_i
N_i^j$~\cite{danon2005}.

\begin{figure}[t!]
\includegraphics[width=6cm, angle=-90]{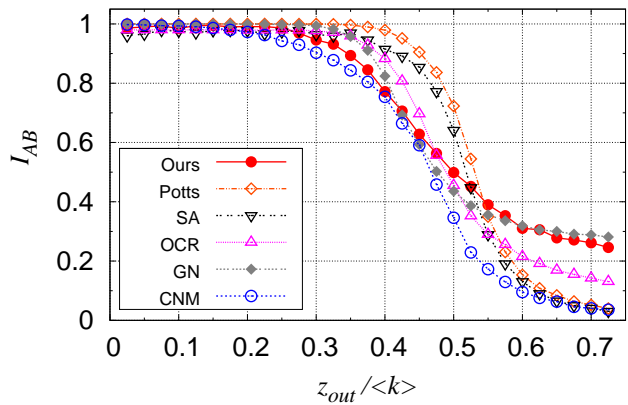}
\caption{(Color online)
The mutual information versus $z_{\rm out}/\langle k\rangle$, the fraction of inter-modular edges per mean degree for the {\it ad hoc} network.
See the text for abbreviations.
}\label{fig_mutual}
\end{figure}

Fig.~\ref{fig_mutual} shows the mutual information measured on the {\it ad hoc} network
as a function of $z_{\rm out}/\langle k \rangle$ for several module detecting algorithms.
The performance of our algorithm is not better than those of the Potts model and the simulated annealing (SA)~\cite{potts,annealing}.
Even though they are better in performance, if we count for their long computation time, then ours may be useful practically.
The performance of opinion changing rate model (OCR) algorithms~\cite{boccaletti}
is somewhat better, however, it requires an extra task of parameter tuning, so that ours is easier to implement.
Since our algorithm shares with the Girvan-Newman (GN) algorithm~\cite{girvan} the idea of clustering based on BC,
the performances of the two algorithms are close to each other. However,
since ours calculates the BC on each edge only once, whereas the GN algorithm does it repeatedly for each disconnected cluster, the computational time can be reduced drastically from $\mathcal{O}(NL^2)$ to $\mathcal{O}(NL)$.
The performance of our algorithm is better than that of the Clauset-Newman-Moore (CNM) algorithm~\cite{clauset}, which runs in $\mathcal{O}(N\ln^2 N)$ for sparse graphs.

\begin{figure}
\includegraphics[width=6cm, angle=-90]{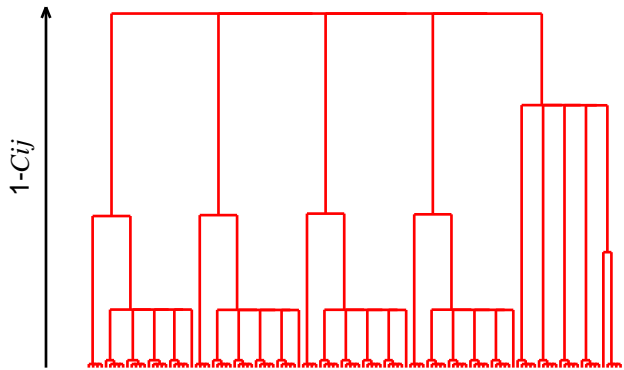}
\caption{(Color online) The dendrogram based on the phase similarity between
connected pairs of vertices for the hierarchical network with three levels.}
\label{fig_dendrogram}
\end{figure}

Secondly, we apply our algorithm to the hierarchical network proposed by Ravasz and Barab\'asi~\cite{ravasz}.
When the number of levels is two, the modules are well selected in a similar way as in Fig.~3 of Ref.~\cite{arenas}.
For the three level case, the dendrogram constructed by our method is shown in Fig.~\ref{fig_dendrogram}.
Here, the hub at the second level is grouped with one of the four identical modules connected to it in the second level.

\begin{figure}
\includegraphics[width=6cm, angle=-90]{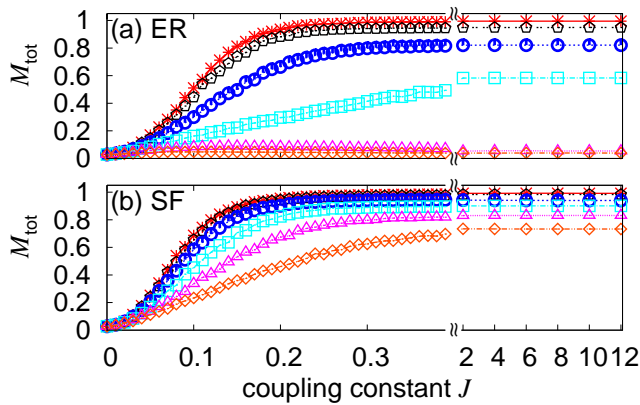}
\caption{(Color online) The order parameter versus the coupling constant $J$
for the ER (a) and the SF network with the degree exponent $3.5$ (b). The data are
for the cases of $\eta=0,0.2,0.4,0.6,0.8$ and $1.0$ from the top.}
\label{fig_M_J}
\end{figure}

\begin{figure}
\includegraphics[width=6cm, angle=-90]{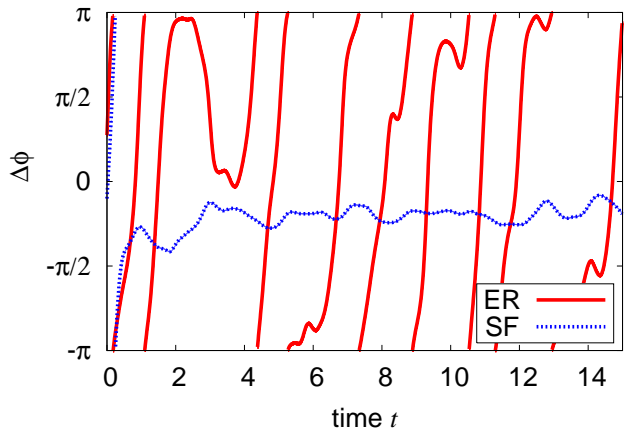}
\caption{(Color online) The phase difference across the edge with the maximum BC.}
\label{fig_phase_diff}
\end{figure}

Thirdly, we apply the gauge KE to Erd\H{o}s-R\'{e}nyi (ER) random networks and scale-free (SF) networks with no modular structures
to see the network structure dependences.
The SF network is generated using the static model~\cite{load}.
The order parameter~(\ref{order}) behaves differently for the two networks.
For the ER network, the saturated value of the order parameter decreases from 1 to 0
as $\eta$ increases from 0 to 1(Fig.~\ref{fig_M_J}(a)).
However, for the SF network, the order parameter does not decrease to 0, but $\approx 0.7$
even if $\eta$ reaches 1(Fig.~\ref{fig_M_J}(b)).
To study the origin of the different behaviors, we measure the phase difference $\Delta\phi$ across the edge with the maximum BC.
In most cases, one end of the edge is the hub.
For the ER network, its change with time is large running from $-\pi$ to $\pi$ as shown in Fig.~\ref{fig_phase_diff}.
For the SF network, it stays around a smaller value in short intervals.
Such difference is rooted from the following.
For the SF network, the hub has large degree,
so that the probability to form a triangle including the miximum BC edge is larger for the SF network than for the ER network,
provided that the mean degree of the system is the same.
Owing to such short loops,
the phase difference across the maximum BC edge is small for the SF network, and large for the ER network.
The overall order parameter is close to zero for the ER network.

In summary, we have introduced a gauge KE in which the gauge term depends on the edge BC.
The gauge term drives the phase difference between the two vertices of an edge from 0 to $\pi$
as the BC across the edge increases.
As a result, the phase difference of two oscillators belonging to different modules is large,
however, it is small across the edges within modules.
Thus, the model generates the phase synchronization within each module, however, it does not globally.
Measuring the phase similarity between two connected oscillators,
we constructed the dendrogram and identified the modules.
Such module detecting method works efficiently.

This work was supported by KOSEF grant Acceleration Research (CNRC) (No.R17-2007-073-01001-0) in SNU and the Korea Research Foundation Grant funded by the Korean Government
(MOEHRD, Basic Research Promotion Fund)(KRF-2007-355-C00030) in KIST.

\end{document}